\documentclass[%
 aip,
 amsmath,amssymb,
preprint,%
]{revtex4-1}

\usepackage{color}
\usepackage{graphicx}
\usepackage{dcolumn}
\usepackage{bm}

\usepackage[utf8]{inputenc}
\usepackage[T1]{fontenc}
\usepackage{mathptmx}
\usepackage{etoolbox}

\makeatletter
\def\@email#1#2{%
 \endgroup
 \patchcmd{\titleblock@produce}
  {\frontmatter@RRAPformat}
  {\frontmatter@RRAPformat{\produce@RRAP{*#1\href{mailto:#2}{#2}}}\frontmatter@RRAPformat}
  {}{}
}%
\makeatother
\begin{document}

\preprint{AIP/123-QED}

\title[ ]{Modulation of Spin Seebeck Effect by Hydrogenation}
\author{K. Ogata}
\affiliation{Department of Integrated Science, University of Tokyo, Meguro, Tokyo 153-8902, Japan}%
\author{T. Kikkawa}
 \affiliation{Department of Applied Physics, University of Tokyo, Bunkyo, Tokyo 113-8656, Japan}
\author{E. Saitoh}
\affiliation{Department of Applied Physics, University of Tokyo, Bunkyo, Tokyo 113-8656, Japan}
\affiliation{Institute for AI and Beyond, University of Tokyo, Bunkyo, Tokyo 113-8656, Japan}
\author{Y. Shiomi}%
 \email{yukishiomi@g.ecc.u-tokyo.ac.jp}
\affiliation{Department of Integrated Science, University of Tokyo, Meguro, Tokyo 153-8902, Japan}%
\affiliation{Department of Basic Science, University of Tokyo, Meguro, Tokyo 153-8902, Japan}

\date{\today}

\begin{abstract}
We demonstrate the modulation of spin Seebeck effect (SSE) by hydrogenation in Pd/YIG bilayers. In the presence of  3\% hydrogen gas, SSE voltage decreases by more than 50 \% from the magnitude observed in pure Ar gas. The modulation of the SSE voltage is reversible, but the recovery of the SSE voltage to the prehydrogenation value takes a few days because of a long time constant of hydrogen desorption. We also demonstrate that the spin Hall magnetoresistance of the identical sample reduces significantly with hydrogen exposure, supporting that the observed modulation of spin current signals originates from hydrogenation of Pd/YIG.       
\end{abstract}

\maketitle
 
Hydrogen is an energy carrier which can be produced by an environmentally clean process and therefore has a positive impact on decarbonization \cite{ma12121973}. To utilize hydrogen as a clean and renewable alternative to carbon-based fuels, hydrogen safety sensors are also critical to assure the development of hydrogen systems \cite{BUTTNER20112462}. Metal-hydride systems have been widely studied for the potential of solid-state hydrogen storage and sensing. In particular, Pd is frequently used as a catalyst for hydrogen dissociation and adsorption. A hydrogen molecule decomposes into independent hydrogen atoms when the molecule approaches a Pd surface due to the strong interaction between Pd and H atoms. As the smallest single atom, a H atom can easily diffuse into the interstices of the Pd lattice and cause lattice expansion. As a result, hydrogen adsorption changes the density of states of Pd near the Fermi energy \cite{PhysRevLett.42.476, PhysRevB.20.3543}, significantly modulating its electrical and optical properties.
\par

The hydrogenation of Pd films also impacts spintronic effects. Pd is known to exhibit a strong spin-orbit coupling and has been used in many spintronic experiments. Magnetic multilayers and super-lattices which include Pd
layers have been of particular interest. It was reported that in Co/Pd bilayers which possess a strong interfacial perpendicular magnetic anisotropy, the magnetic anisotropy and ferromagnetic resonance are reversibly modulated by hydrogen exposure \cite{doi:10.1063/1.4800923, doi:10.1063/1.4893588, PhysRevB.83.094432, OKAMOTO2002313, doi:10.1063/1.4812664, doi:10.1063/1.4800923, doi:10.1063/1.4905463, AKAMARU2015S213}. Efficient hydrogen sensing based on magnetization dynamics was also reported in similar materials \cite{doi:10.1063/1.4800923, doi:10.1063/1.4985241, doi:10.1063/1.4993158}. Moreover, the inverse spin Hall effect (ISHE) induced by spin pumping was successfully modulated by hydrogen exposure in Co/Pd bilayers \cite{8119549, PhysRevB.101.174422}. Absorption of hydrogen gas at 3\% concentration in the Pd layer reduces the ISHE voltage by 20\% \cite{PhysRevB.101.174422}. This decrease in ISHE signals in the presence of hydrogen gas was attributed to the decrease in spin diffusion length due to enhanced scatterings to hydrogen atoms in the Pd layer. After the hydrogen gas is flushed out of the setup, the ISHE voltage returns to the prehydrogenation value; hence the observed effect is reversible.     
\par  

The studies of hydrogen effects on spintronic materials have been carried out for the combination of Pd layers with itinerant magnetic films. For the measurement of ISHE in ferromagnetic/nonmagnetic metallic bilayers, however, it is known that the precise estimation of ISHE voltage is difficult because of spurious spin rectification effects such as anisotropic magnetoresistance and anomalous Hall effect which couple the dynamic magnetization to microwave currents in the ferromagnetic layer \cite{doi:10.7566/JPSJ.86.011003}. Hence magnetic insulators such as Y$_3$Fe$_5$O$_{12}$ (YIG) should be more suitable to investigate hydrogen effects on pure ISHE signals \cite{kajiwara2010transmission}. 
\par

In this letter, we demonstrate the reversible manipulation of the spin Seebeck effect (SSE) by hydrogen exposure in Pd/YIG bilayers. The SSE is the generation of a spin current as a result of a temperature gradient applied across a junction consisting of a magnet and a metal \cite{doi:10.1063/1.3507386, uchida2016thermoelectric}. The spin current injected into a metal can be converted into a voltage by ISHE. Since the ISHE induced by SSE is a transverse thermoelectric effect, it can be employed to realize transverse thermoelectric devices, which could potentially overcome the inherent limitations of conventional thermoelectric devices \cite{D1EE00667C, kikkawa2021observation}. Moreover, SSE is expected to be utilized for flexible heat-flow sensors \cite{kirihara2016flexible}. The manipulation of SSE by hydrogenation demonstrated below may open up new device potentials in spin caloritronics. 
\par

We used epitaxial YIG films with 2 micron thickness grown by liquid phase epitaxy on Gd$_3$Ga$_5$O$_{12}$ (111) substrates. The YIG surfaces were mechanically polished, and then 5-nm-thick Pd films were sputtered at room temperature. The Pd layer is 5 mm long and 0.5 mm wide. For the Pd/YIG bilayers, the SSE measurements in the longitudinal configuration \cite{doi:10.1063/1.3507386} were performed at room temperature using an electromagnet (3470 Electromagnet System, GMW Associates). The bilayer sample was placed between sapphire and copper plates. A 1-k${\rm \Omega}$ resistive heater was attached to the upper sapphire plate and the lower copper plate is a heat sink. To facilitate hydrogenation of the Pd layer, a breathable tape (TBAT-252, TRUSCO) was inserted between the upper plate and the sample [Fig. 1(a)]. The temperature gradient is generated by applying an electric current to the heater. Two-pairs of leads were attached to the Pd layer to measure not only the SSE but also the spin Hall magnetoresistance (SMR) \cite{PhysRevLett.110.206601} in the same setup. The distance between the voltage terminals is 2.5 mm. The thermoelectric voltage due to the ISHE induced by SSE was monitored with a Keithley 2182A nanovoltmeter. SMR was measured by lockin detection using Anfatec USB Lockin Amplifier 250; the frequency and amplitude of ac electric current are 111 Hz and 0.8 mA. The sample was loaded into a small chamber to control the atmosphere. For hydrogenation measurements, the samples were first measured in pure Ar gas ($>$99.9999 vol.\%) at atmospheric pressure followed by a 3\%/97\% H$_2$/Ar gas mixture. Before the measurements in Ar-H$_2$ gas, we waited 20-40 minutes for the Pd layer to be completely hydrogenated \cite{PhysRevB.101.174422, lederman2004magnetooptic} after the chamber was filled with 1 atm Ar-H$_2$ gas.   
\par

\begin{figure}[t]
\begin{center}
\includegraphics[width=10cm]{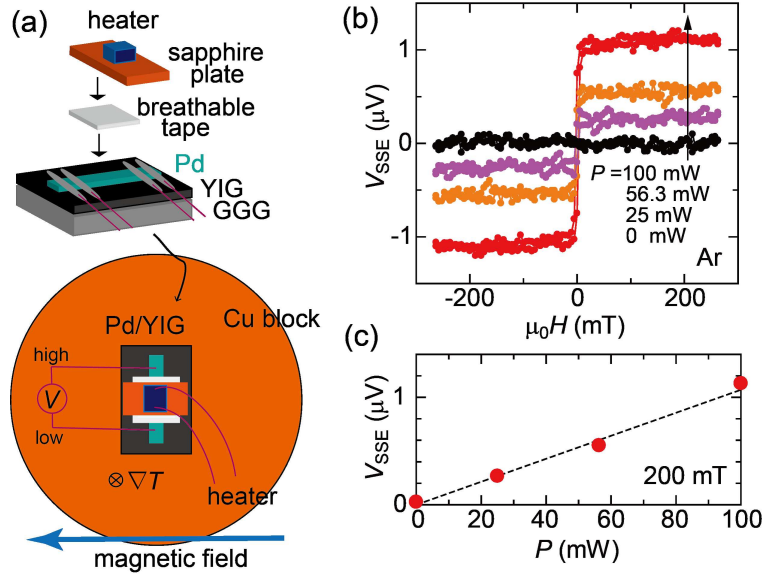}
\caption{(a) Measurement setup of the SSE. A breathable tape was inserted between Pd/YIG and the heater part to facilitate hydrogen absorption and desorption in the Pd film. (b) Magnetic field ($H$) dependence of the SSE voltage ($V_{SSE}$) measured in 1 atm of Ar. The heater power ($P$) was changed from 0 mW to 100 mW. (c) Heater power ($P$) dependence of the SSE voltage ($V_{SSE}$) at 200 mT. The raw data is shown in (b). } 
\label{fig1}
\end{center}
\end{figure}

First, we measured SSE voltage $V_{SSE}$ of Pd/YIG in Ar atmosphere. Figure 1(b) shows the magnetic-field ($H$) dependence of $V_{SSE}$ measured at several heater power levels. Here symmetric components of the output voltage with respect to $H$ are subtracted and antisymmetric components are plotted; note that the symmetric components which are not to be attributed to the effect under study are almost independent of $H$ in our measurements. When the heater is off, $V_{SSE}$ is almost zero in the entire $H$ range in Fig. 1(b). As the heater power $P$ increases from zero, the clear SSE signals appear and their magnitudes increase with $P$. The sign of $V_{SSE}$ is the same as that reported for Pt/YIG \cite{doi:10.1063/1.3507386}. The saturated magnitude of $V_{SSE}$ is plotted against $P$ in Fig. 1(c). The $V_{SSE}$ magnitude increases linearly with the heater power, which indicates that $V_{SSE}$ is proportional to temperature gradient generated across the Pd/YIG junction. The temperature difference $\Delta T$ generated at $P=100$ mW is estimated to be $\sim$1.5 K (see Fig. S1 in Supplementary Material). 
\par  

\begin{figure}[t]
\begin{center}
\includegraphics[width=10cm]{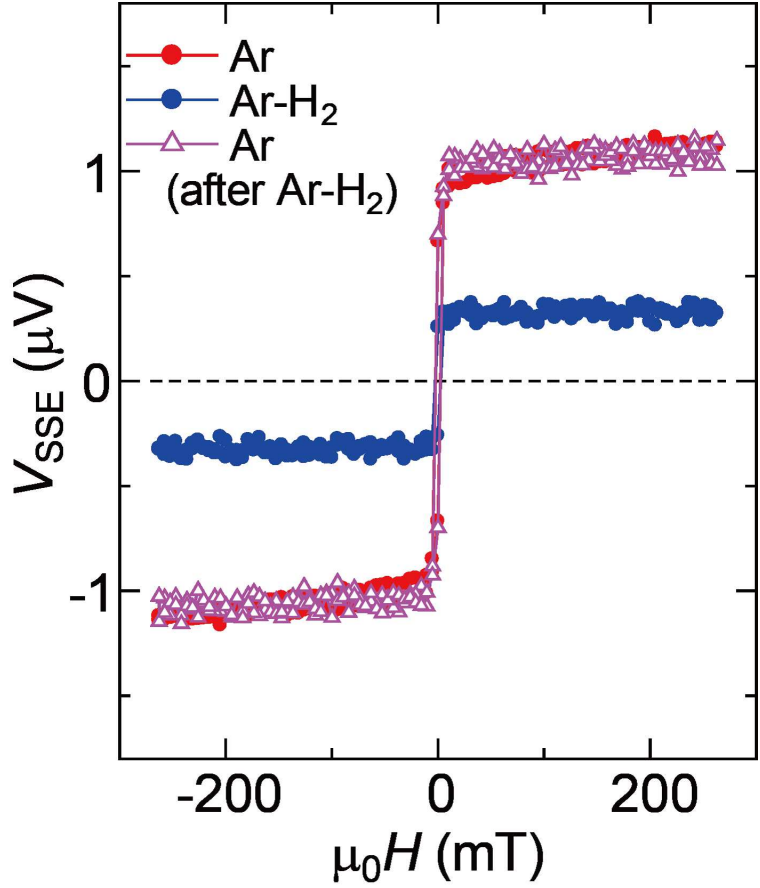}
\caption{Magnetic field ($H$) dependence of the SSE voltage ($V_{SSE}$) measured before hydrogenation, during exposure to hydrogen gas, and after the hydrogen gas is flushed out of the setup. The heater power is kept at $100$ mW.} 
\label{fig2}
\end{center}
\end{figure}

Next, the effect of exposing the 3\% H$_2$ mixture on the Pd/YIG sample is investigated in Fig. 2 (see also Fig. S2 in Supplementary Material for additional experimental results). Here the heater power $P$ is kept constant at 100 mW during the series of measurements. After the initial SSE measurement in pure Ar gas at atmospheric pressure already shown in Figs. 1(b) and 1(c), the sample chamber was filled with H$_2$ 3\% H$_2$/Ar mixture and the SSE measurement was performed. As shown in Fig. 2, the magnitude of $V_{SSE}$ is found to be reduced by more than 50\% in the presence of H$_2$ gas. After completing the SSE measurement in Ar-H$_2$ atmosphere, the sample was then remeasured in pure Ar. The $V_{SSE}$ magnitude returned to the pristine value as shown in Fig. 2. Note that this data was taken 2.5 days after the chamber was refilled with Ar. The observed decrease in the SSE signal is safely ascribed to the presence of hydrogen in Pd/YIG, and importantly, the change is reversible.   
\par

It is well known \cite{PhysRevB.101.174422, doi:10.1063/1.4986214} that upon hydrogenation, Pd thin films undergo two stages of lattice expansion depending on the hydrogen gas concentration. For light concentration levels up to 2-3 \%, the lattice constant grows by approximately 1\% in the out-of-plane direction only. This expansion is reversible. In the second stage, the lattice constant grows by up to 4\% in both out-of-plane and in-plane directions. These changes are irreversible, causing structural changes to the Pd lattice. In our SSE measurements under 3\% hydrogen gas, the sample should undergo the first stage of lattice expansion and the SSE is thereby reversible. Note that we confirmed by x-ray diffraction that the Pd films are (111) oriented as in the literature \cite{doi:10.1063/1.4986214} (see Supplementary Material).
\par

\begin{figure}[t]
\begin{center}
\includegraphics[width=10cm]{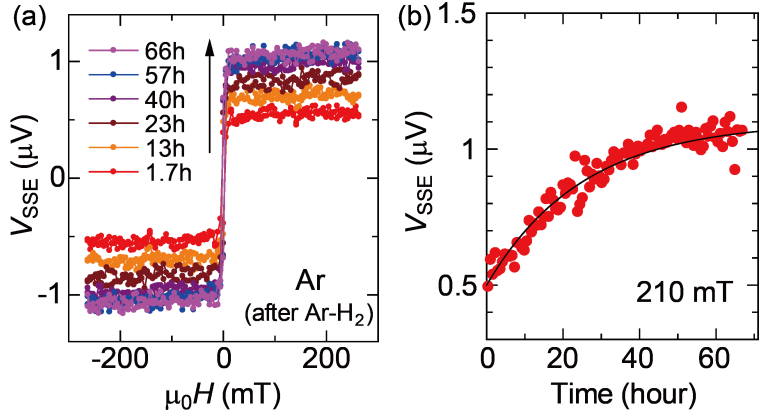}
\caption{(a) Magnetic field ($H$) dependence of the SSE voltage ($V_{SSE}$) measured 1.7-66 hours after Ar gas is refilled in the measurement chamber. The heater power is kept at $100$ mW. (b) Time dependence of the SSE voltage ($V_{SSE}$) at 210 mT measured after Ar gas is refilled in the measurement chamber. The selected raw data is shown in (a). The black curve is a fit to the experimental data (see text).   } 
\label{fig3}
\end{center}
\end{figure}

Though the modulation of SSE by hydrogen absorption/desorption is reversible, the recovery time of the SSE signal due to hydrogen desorption is as long as 2.5 days. It was reported that the hydrogen desorption takes a long time in contrast to the quick hydrogen absorption \cite{doi:10.1063/1.4812664}, and the response time depends significantly on materials. The time for hydrogen desorption is typically at most several tens of minutes for Co/Pd \cite{doi:10.1063/1.4800923, doi:10.1063/1.4893588, PhysRevB.83.094432, OKAMOTO2002313, doi:10.1063/1.4812664, doi:10.1063/1.4800923, doi:10.1063/1.4905463, AKAMARU2015S213}, while the completion of the entire desorption requires at least a few days at $10^{-3}$ mbar for FePd alloys \cite{doi:10.1063/1.5142625}. Our Pd/YIG also includes Fe and Pd, and the situation looks similar to FePd alloys.
\par

We then take a closer look on the dehydrogenation process by the time dependent measurement of SSE in Fig. 3. Figure 3(a) shows $V_{SSE}$ curves measured at different times after the measurement chamber is refilled with pure Ar gas. The $V_{SSE}$ magnitude is approximately 0.5 ${\rm \mu}$V just after the gas is replaced with Ar, and increases monotonically with time. After 50 hours, the $V_{SSE}$ magnitude is almost saturated at $\sim 1$ ${\rm \mu V}$.
\par

The time dependence of $V_{SSE}$ at 210 mT is plotted in Fig. 3(b). The $V_{SSE}$ magnitude increases monotonically with time, as already shown in Fig. 3(a). We fit the experimental data by a standard relaxation function: $V_{SSE} \propto 1- e^{-t/\tau}$, where $t$ is the measurement time and $\tau$ is a time constant of hydrogen desorption. The fitting curve matches the experimental data very well, meaning that the hydrogen desorption follows an exponential function. The same function was adopted for the hydrogenation effect on magneto-optical effects in  
Pd/Co/Pd films \cite{doi:10.1063/1.4812664}. The fit in Fig. 3(b) yields $\tau \approx 25$ hour. Such a long time constant was not observed in the spin pumping measurement for Pd/Co bilayers \cite{PhysRevB.101.174422}.   
\par

In contrast to the spin pumping measurements, the attachment of the heater to the Pd surface is required in the SSE measurements, which may adversely affect the absorption/desorption of hydrogen because of small numbers of exposed surface atoms. To confirm that spin current signals in the Pd layer is indeed modulated by hydrogenation, we also perform the measurement of SMR (spin Hall magnetoresistance) for the same sample in the same setup. The SMR is a magnetoresistance effect related to a nonequilibrium proximity effect caused by the simultaneous action of the SHE and ISHE \cite{PhysRevLett.110.206601, PhysRevB.87.144411}; the absorption/reflection of spin current at the ferromagnet/metal interface results in magnetoresistance, since the spin-dependent scattering at the metal/ferromagnet interface depends on the angle between the polarization of spin Hall current and the magnetization of the attached magnetic layer. The experimental setup is illustrated in the inset to Fig. 4(a). Magnetic field is applied perpendicular to the electric-current direction in the film plane.
\par

\begin{figure}[t]
\begin{center}
\includegraphics[width=10cm]{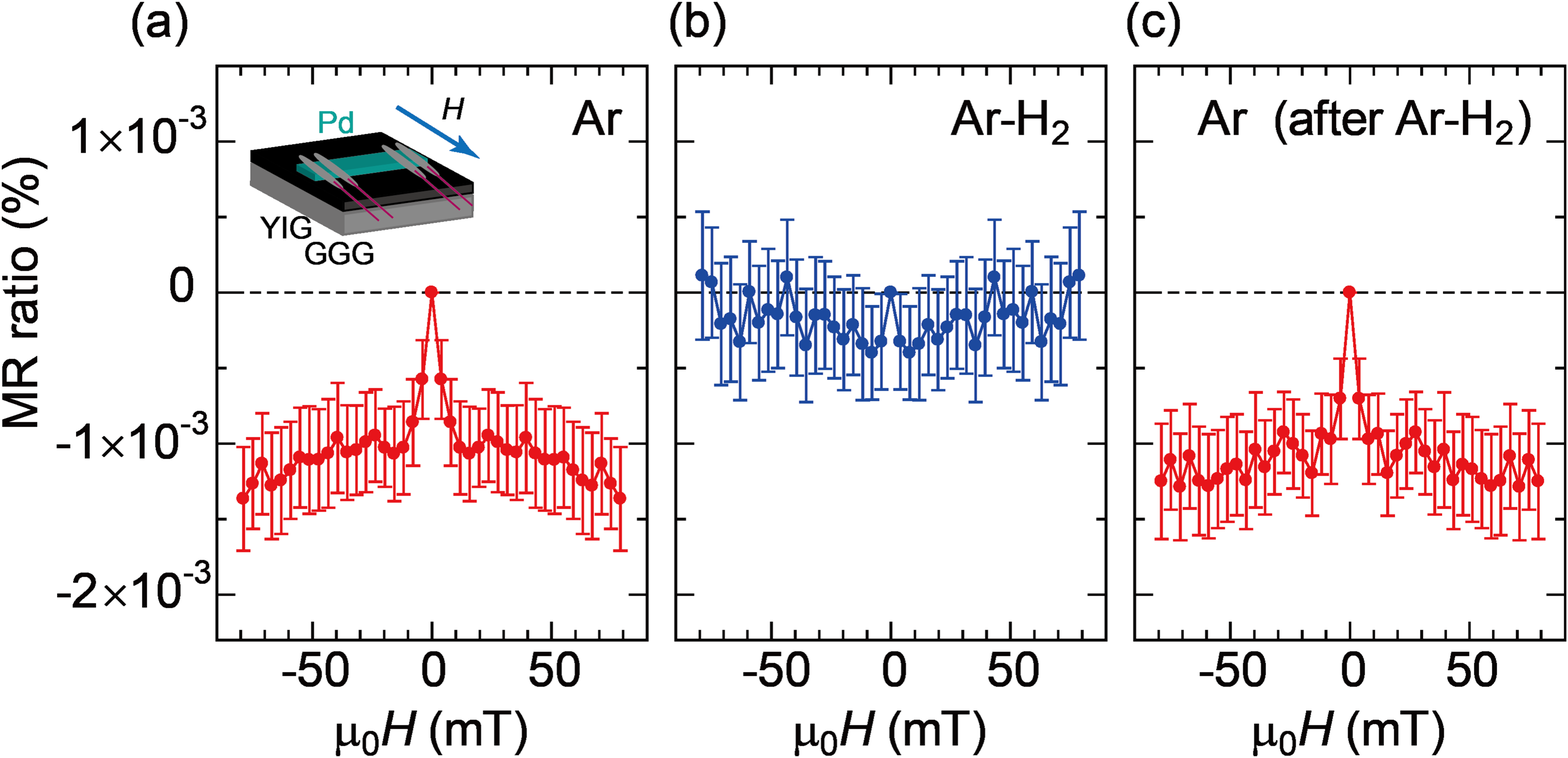}
\caption{Magnetic field ($H$) dependence of the magnetoresistance (MR) ratio ($\rho(H)/\rho(H=0) -1$) measured before hydrogenation (a), during exposure to hydrogen gas (b), and after the hydrogen gas is flushed out of the setup (c).   } 
\label{fig4}
\end{center}
\end{figure}

Figure 4 shows the hydrogen effects on SMR in the Pd/YIG bilayer. Here, since the size of SMR is very small, the magnetoresistance measurements were repeated several times and averaged. The error bars stand for the standard errors. Before hydrogenation [Fig. 4(a)], a negative magnetoresistance effect is observed. The magnetic-field dependence of resistance change follows the magnetization process of the YIG layer, consistent with the SMR \cite{PhysRevLett.110.206601}. The size of SMR is about $1 \times 10^{-3}$\%. This magnitude is about ten times smaller than that in Pt/YIG \cite{PhysRevLett.110.206601}. A small SMR of about 10\% compared to Pt/YIG was also reported in the literature \cite{PhysRevB.98.224424}.        
\par

During the exposure to 3\% hydrogen gas, the SMR magnitude decreases significantly as shown in Fig. 4(b). Although quantitative analysis is difficult because of the large error bars, the suppression of SMR ratio by hydrogenation looks more than 50\%, consistent with the modulation in SSE voltages (Figs. 2 and 3). After the hydrogen gas is flushed out of the chamber and pure Ar gas is refilled, we confirmed that the size of SMR returns to the initial value [Fig. 4(c)].    
\par

An important finding in the SMR measurement is that the SMR ratio has already returned to its original value 30 minutes after refilling Ar gas. Namely, the time constant of hydrogen desorption in the SMR measurement is much shorter than that in the SSE measurement. Since both the measurements were performed for the same sample in the same setup, the long time constant of hydrogen desorption in the SSE measurement cannot be attributed to impurities/defects in the Pd layer, surface oxidation, surface morphology \cite{doi:10.1063/1.4812664}, or moisture which may trap hydrogen atoms and hinder the hydrogen desorption  \cite{JOSE20106804, KISHORE2005234}. 
\par

In our measurements of SSE and SMR, spin current signals are significantly suppressed by hydrogen exposure as shown in Figs. 2-4. The decrease in the spin Hall signals with hydrogen exposure is consistent with the previous spin pumping measurements for Co/Pd \cite{8119549, PhysRevB.101.174422}. Scatterings of conduction electrons to hydrogen atoms in the Pd layer decrease the spin diffusion length due to the enhanced Elliot-Yafet relaxation mechanism, and result in the decrease in spin-pumping signals \cite{PhysRevB.101.174422}. This mechanism should be also applicable to SSE and SMR. Since the theory has shown that both effects depend on spin diffusion length and spin Hall angle of the Pd layer \cite{uchida2016thermoelectric, PhysRevB.87.144411}, the signal variation by hydrogenation can be attributed to the decrease in the spin diffusion length \cite{PhysRevB.101.174422}. On the other hand, it is notable that magneto-optical Kerr signals are enhanced by hydrogenation in Co/Pd bilayers \cite{doi:10.1063/1.4905463, doi:10.1063/1.4812664}, in contrast to the decrease in the spin current signals \cite{8119549, PhysRevB.101.174422}. In transport measurements such as (inverse) spin Hall effects, enhanced electron scatterings due to interstitial hydrogen impurities are likely to play a dominant role in the hydrogenation effect. The significant scattering effect due to hydrogen atoms is also evidenced by the reduction of the anomalous Hall signal in hydrogenated Co$_{x}$Pd$_{1-x}$ films \cite{doi:10.1063/1.4985241}. 
\par

The decrease in the $V_{SSE}$ magnitude ($>50$\%) by hydrogen exposure is noticeably greater than the change in ISHE signals reported in the spin pumping measurements \cite{PhysRevB.101.174422}; the decrease in the spin-pumping voltage in H$_2$/Ar mixture with 3\% of hydrogen was only 20\%. The larger signal change in our results suggests that there may be other factors for the reduction of $V_{SSE}$ besides hydrogenation of the Pd layer. The first possibility is imperfect separation of ISHE signals from spin rectification effects in metallic Co/Pd bilayers \cite{8119549, PhysRevB.101.174422}. Another possible origin is different interfacial stresses to the Pd layer between YIG and Co. It is known that electrical resistivity of single-layer Pd grown on Si substrates increases upon hydorogenation, while it tends to decrease for bilayer cases \cite{PhysRevB.101.174422} because of the interfacial compressive stress from the underlying layer. The interfacial stress can also affect the interface spin mixing conductance, modulating the injection efficiency of spin currents. Note that the resistivity of the Pd film on YIG decreases by hydrogenation, but the change in resistivity is as small as 1\% (Fig. S3 in Supplementary Material), which cannot explain the large variation ($>$ 50\%) of SSE voltage by hydrogen exposure. 
\par

Moreover, since the SSE also depends on bulk spin transport in the YIG layer \cite{PhysRevB.95.174401, PhysRevX.6.031012} in contrast to the spin pumping, hydrogen effects on YIG may contribute to the significant reduction in the $V_{SSE}$ magnitude. Hydrogen diffusion in YIG was indeed reported for annealed samples in H$_2$ atmosphere \cite{doi:10.1063/1.333600, MILANI198573, doi:10.1063/1.337383}. The hydrogen diffusion in the YIG layer can suppress the magnon and phonon transport, which should reduce the $V_{SSE}$. Also the interface spin-exchange coupling can be weakened by hydrogen around the interface, leading to the decrease in the interface spin-injection efficiency.
\par

The presence of hydrogen effects on the YIG layer is also suggested by the different recovery time constants between SSE and SMR. Our measurements in Figs. 3 and 4 showed that the time constant for the signal recovery of SSE is much longer than that of SMR.  The different recovery time constants are attributable to different bulk sensitivity of these effects. In SMR, spin-dependent scattering at the Pd/YIG interface is essential. In contrast, bulk thermal spin current also plays an important role in the SSE voltage \cite{PhysRevB.95.174401, PhysRevX.6.031012} in addition to the interfacial spin coupling. Bulk properties of magnetic materials such as bulk magnetization, thermal conductivity, and magnon transport coefficient contribute to the SSE signals, but not to SMR or spin pumping. Also in the case of magneto-optical effects of Pd/Co frequently studied before for hydrogenation effects, the variation of perpendicular magnetic anisotropy originates from interface effects \cite{doi:10.1063/1.4800923, doi:10.1063/1.4812664}. Hence the SSE is a rare spintronic phenomenon that depends not only on interface properties but also on bulk properties of magnons and phonons in the magnetic layer. Hydrogen effects on YIG could be related to the reduction of $V_{SSE}$ and also the long time constant for hydrogen desorption in SSE. 
\par

In conclusion, we experimentally demonstrated the reversible modulation of SSE and SMR by hydrogenation in Pd/YIG bilayers. Absorption of hydrogen results in the decrease in both SSE and SMR signals by more than 50\%. Enhanced scatterings of conduction electrons to hydrogen atoms in the Pd layer are partly responsible for the decrease in the spin-current signals, as reported in the previous spin pumping experiments. The modulation of SSE voltage is reversible, but the time constant for the signal recovery is longer than 2 days. The long time constant for hydrogen desorption in the SSE measurement is in contrast to the case of SMR, in which the SMR ratio already returned to the prehydrogenation value 30 minutes after the chamber was refilled with pure Ar. We speculate that the significant decrease in the SSE magnitude by hydrogen exposure and the long time constant for hydrogen desorption in SSE are related to the hydrogen modulation of bulk properties of the YIG layer, since the SSE depends not only on interfacial spin couplings but also on bulk properties of the magnetic layer. We hope that the present results will stimulate further research on hydrogen effects on Pd films grown on insulating magnetic oxides.    
\par

See the supplementary material for additional SSE data, x-ray diffraction data, and resistivity change by hydrogen exposure.
\par

We thank Y. Miyazaki for the experimental help of sample preparation and Dr. T. Yokouchi for the fruitful discussion. This research was supported by JST CREST (JPMJCR20C1 and JPMJCR20T2), Institute for AI and Beyond of the University of Tokyo, and JSPS KAKENHI Grant Numbers JP20H05153, JP20H02599, JP20H04631, JP21K18890, JP19H05600, and JP19H02424. 
\par


The data that support the findings of this study are available from the corresponding author upon reasonable request.




\bibliography{bib-main} 
\bibliographystyle{unsrt}

\end{document}